\documentclass{article}

\PassOptionsToPackage{authoryear,round}{natbib} % author–year
 \usepackage[preprint]{neurips_2025}
 \workshoptitle{Data on the Brain \& Mind}
 \usepackage{graphicx} 
 \usepackage{amsmath}

\workshoptitle{Foundation Models for the Brain and Body}

\usepackage[utf8]{inputenc} % allow utf-8 input
\usepackage[T1]{fontenc}    % use 8-bit T1 fonts
\usepackage{hyperref}       % hyperlinks
\usepackage{url}            % simple URL typesetting
\usepackage{booktabs}       % professional-quality tables
\usepackage{amsfonts}       % blackboard math symbols
\usepackage{nicefrac}       % compact symbols for 1/2, etc.
\usepackage{microtype}      % microtypography
\usepackage{xcolor}         % colors

\newcommand{\repourl}{\url{https://anonymous.4open.science/r/GBM-8E94/}}
\newcommand{\demourl}{\url{https://virtbrain.samfv.systems}}

\title{A Sensing Whole Brain Zebrafish Foundation Model for Neuron Dynamics and Behavior}

\author{%
  Sam Fatehmanesh Vegas \\
  California Institute of Technology\\
  \texttt{sfatehma@caltech.edu} \\
  \And
  Matt Thomson \\
  California Institute of Technology \\
  \texttt{mthomson@caltech.edu} \\
  \AND
  James Gornet \\
  California Institute of Technology \\
  \texttt{jgornet@caltech.edu} \\
  \And
  David Prober \\
  California Institute of Technology \\
  \texttt{dprober@caltech.edu} \\
}

\begin{document}

\maketitle

\begin{abstract}
Neural dynamics underlie behaviors from memory to sleep, yet identifying mechanisms for higher-order phenomena (e.g., social interaction) is experimentally challenging. Existing whole-brain models often fail to scale to single-neuron resolution, omit behavioral readouts, or rely on PCA/conv pipelines that miss long-range, non-linear interactions. We introduce a sparse-attention whole-brain foundation model (SBM) for larval zebrafish that forecasts neuron spike probabilities \emph{conditioned on sensory stimuli} and links brain state to behavior. SBM factorizes attention across neurons and along time, enabling whole-brain scale and interpretability. On a held-out subject, it achieves mean absolute error $<\!0.02$ with calibrated predictions and stable autoregressive rollouts. Coupled to a permutation-invariant behavior head, SBM enables gradient-based synthesis of neural patterns that elicit target behaviors. This framework supports rapid, behavior-grounded exploration of complex neural phenomena.

\end{abstract}

\section{Introduction}

Predicting large-scale neural dynamics at single-neuron resolution is essential for linking brain activity to behavior and for running rapid in silico experiments that can guide in vivo work \citep{Ahrens2013,Naumann2016}. Whole-brain light-sheet imaging and modern calcium indicators now make it possible to record brain-wide activity at cellular resolution in larval zebrafish during visually guided behavior \citep{Ahrens2013,Vladimirov2014,Dana2019}. Yet current modeling approaches struggle to meet five goals at once: accurate next-step prediction, fidelity to the distribution of brain states, scalability to whole brains, behavioral coverage, and interpretability. PCA pipelines compress activity into low-dimensional embeddings that discard neuron-level structure, limiting connectomic or functional interpretation and often failing at whole-brain scale \citep{Jolliffe2016}. Convolutional video architectures such as U-Nets and their 3D variants emphasize local receptive fields and demand substantial compute when extended to high-resolution spatiotemporal volumes, and popular video diffusion systems continue to rely on convolutional backbones, underscoring the computational burden for long-range interactions \citep{Ronneberger2015,Cicek2016,Ho2022VDM,Ho2022ImagenVideo}. Recent forecasting benchmarks highlight the opportunity for foundation-model style approaches in neural video but do not yet provide single-neuron interpretability together with behavior-level readouts \citep{Immer2025,Duan2025POCO}. 

To address this gap we introduce the Sparse Brain Model (SBM), which factorizes attention along space and time with two modules. A dynamic connectome layer applies time-independent self-attention across neuron tokens to expose neuron-neuron influences. A temporal neuron layer applies causal self-attention within each neuron’s history, enabling long-range temporal dependence without inter-neuron confounds. We derive neuron-level spike probabilities from DF/F traces using a causal spike-inference pipeline so that the model learns on spiking statistics rather than raw fluorescence \citep{Rupprecht2021}. The attention mechanism provides global, distance-independent interactions and efficient parallelism, and rotary position embeddings provide a compact way to encode relative position for temporal attention \citep{Vaswani2017,Su2021}. To connect brain state to action, a Peripheral Neural Model (PNM) reads either ground truth or predicted brain states and maps them to behavior, enabling both behavioral prediction and optimization of neural inputs that elicit target actions \citep{Naumann2016}. 
We find that sparse attention at neuron resolution yields accurate next-step predictions with strong calibration, lower error with longer context, and sublinear error growth during autoregressive rollout. Predicted and true brain states occupy similar low-dimensional manifolds in PCA and UMAP spaces, supporting distributional fidelity rather than only pointwise accuracy \citep{Jolliffe2016,McInnes2018}. Coupled with the PNM, the system predicts fish behaviors from short histories of brain state and permits gradient-based synthesis of neural patterns that expand the reachable behavioral space beyond random stimulation, providing testable hypotheses for mechanisms that link brain-wide dynamics to action \citep{Naumann2016}. We train and validate all models on the publicly released larval zebrafish whole-brain calcium-imaging dataset of \citet{Chen2018}.

An interactive web demo for region- and neuron-level simulated optogenetic experiments utilizing the whole brain foundation model is available at \demourl. 
Model code, data loading, and training scripts available at \repourl.

\begin{figure}
  \centering
  \includegraphics[width=1\linewidth]{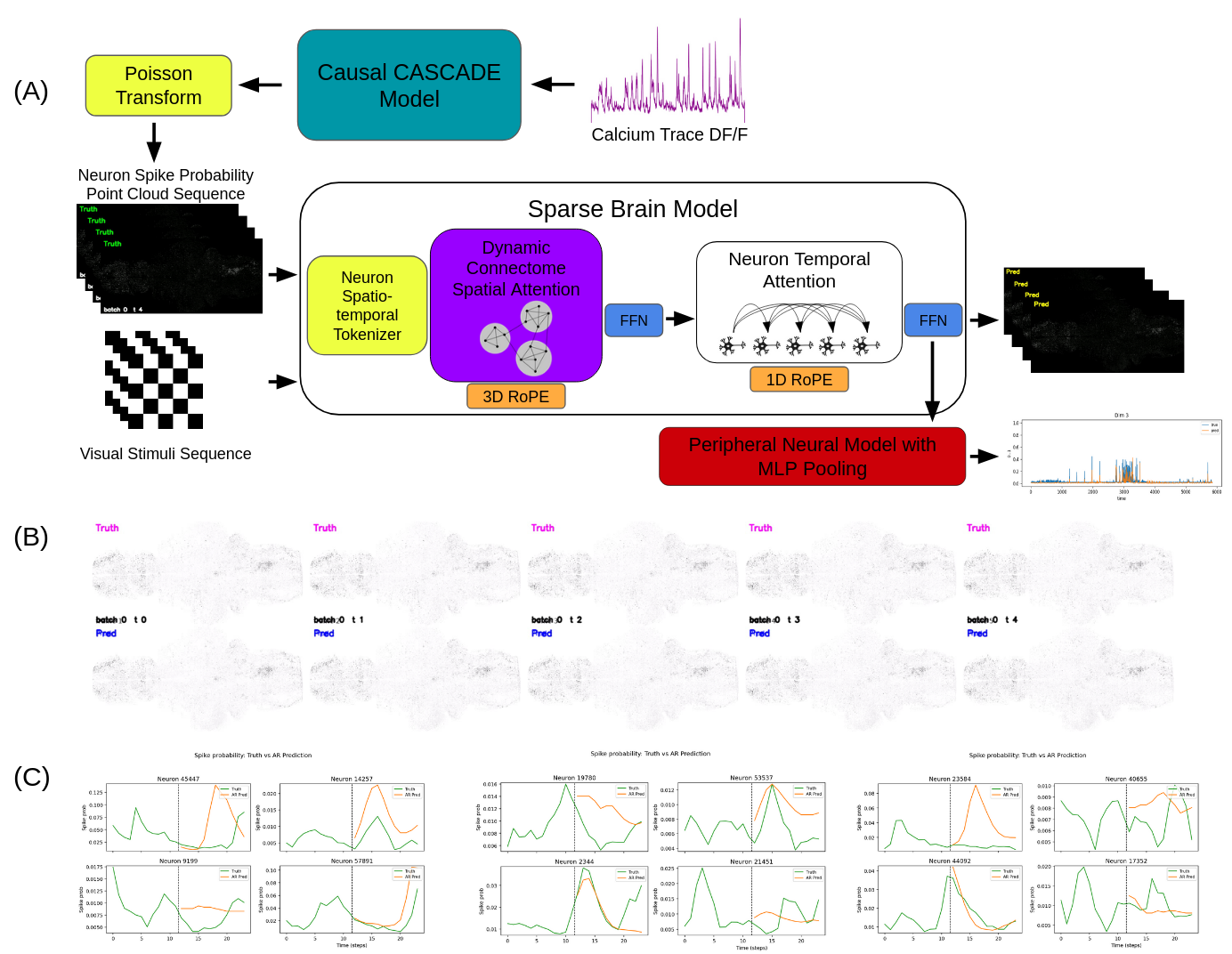}
  \caption{Sparse Attention Architecture enables whole-brain foundation models. (A) This panel illustrates the model architecture and data-processing pipeline. A causal CASCADE model \citep{Rupprecht2021} is applied to DF/F neuron calcium traces to estimate spike rates, which are then converted to spike probabilities under a Poisson assumption. The Sparse Brain Model (SBM) comprises two key layers: a dynamic connectome layer that infers functional neural clusters and applies time-independent self-attention across neuron tokens, and a temporal neuron layer that applies causal self-attention to each neuron’s token sequence independently of other neurons. (B) Comparison of ground-truth and next-step predictions from the SBM, visualized as neuron spike-probability point clouds voxelized to a 512×256 grid and mean-pooled across the z-axis. (C) Comparison of ground-truth spike-probability traces with autoregressive predictions generated using a 4-second sliding window.}
  \label{fig:arch}
\end{figure}

\begin{figure}
  \centering
  \includegraphics[width=1\linewidth]{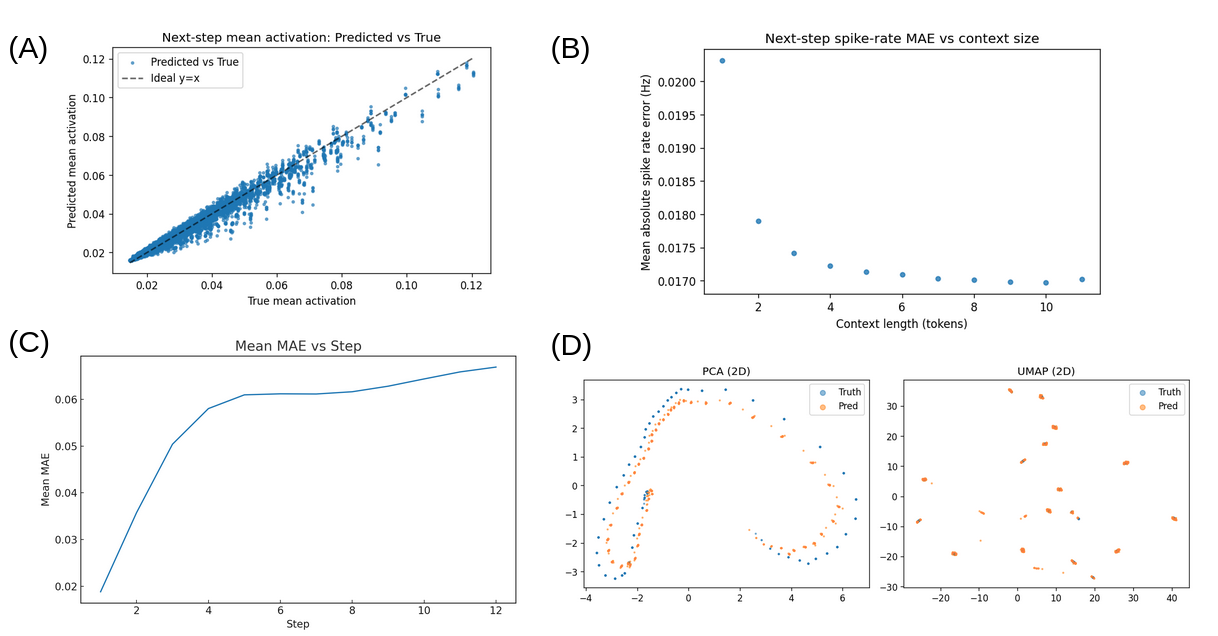}
  \caption{Sparse Brain Models produce more accurate next-step predictions with longer context and preserve low error during autoregression. (A) The SBM is well calibrated, with predicted mean neuron spike probabilities highly correlated with ground-truth means. (B) Prediction error decreases as the history size increases, and during autoregression the error accumulates sublinearly with respect to the number of steps when using a 4-second (12 time-point) sliding window. (C) During next-step prediction, greater temporal context yields significantly lower error.(D) In PCA and UMAP embeddings, the distributions of next-step predictions and ground-truth brain states are highly correlated, indicating that the SBM preserves the brain’s original activity distribution and manifold.}
  \label{fig:metrics}
\end{figure}

%\section{Results}
%\label{gen_inst}

% Model architecture and training
\section{Model architecture and training}
\label{sec:model}

\paragraph{Problem setup.}
Let $x_{t,n}\in[0,1]$ denote the spike \emph{probability} of neuron $n\in\{1,\dots,N\}$ at time $t\in\{1,\dots,T\}$, inferred \emph{causally} from calcium traces via CASCADE \citep{Rupprecht2021}. Let $p_n\in\mathbb{R}^3$ be the soma location and $s_t\in\mathbb{R}^{d_s}$ encode exogenous input (stimuli/task). The task is next-step forecasting: given a context window $\mathcal{C}_t=\{(x_{t-\tau:t-1,\cdot},\, s_{t-\tau:t-1})\}$ of length $\tau$, predict $x_{t,\cdot}$; during rollout we iterate this autoregressively.

\subsection{Sparse Brain Model (SBM)}
\label{sec:sbm}

SBM factorizes spatiotemporal reasoning into two attention modules per block: (\emph{i}) a \emph{dynamic connectome layer} that attends across neurons (plus a stimulus token) within each time step, and (\emph{ii}) a \emph{temporal neuron layer} that applies \emph{causal} attention along the history of each neuron independently. This preserves single-cell interpretability while scaling to whole brains.

\paragraph{Tokenization and embeddings.}
At time $t$ we form $N$ neuron tokens and one stimulus token. Neuron $n$ is embedded from its scalar activity and 3D position using RMSNorm \citep{Zhang2019RMSNorm}:
\begin{equation}
\label{eq:neuron-embed}
h^{(0)}_{t,n}=\mathrm{RMSNorm}\!\big(W_{\text{neuron}}\,[x_{t,n};\,p_n]+b_{\text{neuron}}\big)\in\mathbb{R}^{d},
\end{equation}
and the stimulus token is
\begin{equation}
\label{eq:stim-embed}
h^{(0)}_{t,\text{stim}}=\mathrm{RMSNorm}\!\big(W_{\text{stim}}\, s_t + b_{\text{stim}}\big)\in\mathbb{R}^{d}.
\end{equation}
We stack $H^{(0)}_t=[h^{(0)}_{t,1},\dots,h^{(0)}_{t,N},h^{(0)}_{t,\text{stim}}]\in\mathbb{R}^{(N+1)\times d}$. Variable neuron counts are padded to $N_{\max}$ with mask $m\in\{0,1\}^{N_{\max}}$ respected by all attention ops.

\paragraph{Block structure.}
Each of the $L$ blocks applies spatial then temporal attention with residuals and normalization:
\begin{align}
\widetilde{H}^{(\ell)}_t &= \mathrm{SpatialAttn}\!\big(\mathrm{RMSNorm}(H^{(\ell-1)}_t),\,P\big) + H^{(\ell-1)}_t,\quad P=\{p_n\}_{n=1}^{N}, \label{eq:spatial}\\
Z^{(\ell)}_{\cdot,n} &= \mathrm{TemporalAttn}\!\big(\mathrm{RMSNorm}(\widetilde{H}^{(\ell)}_{\cdot,n}),\,\text{causal}\big) + \widetilde{H}^{(\ell)}_{\cdot,n}, \label{eq:temporal}
\end{align}
where $H^{(\ell)}_t$ stacks $\{Z^{(\ell)}_{t,n}\}_{n=1}^{N}$ and drops the stimulus token after the spatial layer.

\paragraph{Dynamic connectome layer (spatial attention).}
Within each time slice we attend \emph{across} neurons (and the stimulus token). 3D geometry is injected via directional rotary encodings (spatial RoPE): project $p_n$ onto random unit directions at log-spaced frequencies and add to Q/K streams \citep{Su2021}. To scale to $N\!\sim\!10^5$, we use \emph{routing} to partition tokens into $k$ balanced clusters of target size $w\!\ll\!N$; each centroid selects top-$w$ tokens (multi-membership allowed), attention is computed per cluster with variable-length FlashAttention, and outputs are scatter-added to the original order and averaged over duplicates. This reduces cost from $\mathcal{O}(N^2)$ to $\mathcal{O}(k\,w^2)$ with $k\!\approx\!\lceil N/w\rceil$, i.e., effectively linear in $N$ for fixed $w$ \citep{Roy2020Routing,Dao2023FlashAttn2}.

\paragraph{Temporal neuron layer (causal attention).}
For each neuron $n$, we apply causal multi-head attention along time with standard RoPE over indices:
\begin{equation}
\mathrm{TemporalAttn}(Q,K,V)=\mathrm{MHA}\big(\mathrm{RoPE}_t(Q),\,\mathrm{RoPE}_t(K),\,V;\ \text{causal}\big), \label{eq:temporal-attn}
\end{equation}
batching neurons as independent sequences of length $\tau$ (masked rows skipped). This captures long-range temporal dependence per cell without quadratic cross-neuron time interactions.

\paragraph{Decoder and training loss.}
After $L$ blocks we output a logit per neuron and time,
\begin{equation}
z_{t,n}=w_\text{dec}^\top \mathrm{RMSNorm}\!\big(Z^{(L)}_{t,n}\big)+b_\text{dec},\qquad \hat{x}_{t,n}=\sigma(z_{t,n}),
\end{equation}
and train with binary cross-entropy over unpadded tokens:
\begin{equation}
\label{eq:bce}
\mathcal{L}_{\text{BCE}} \;=\; - \frac{1}{|\mathcal{I}|}\sum_{(t,n)\in \mathcal{I}}
\Big[x_{t,n}\log \sigma(z_{t,n}) + (1-x_{t,n})\log(1-\sigma(z_{t,n}))\Big].
\end{equation}

\paragraph{Autoregressive rollout.}
Given $(X_{1:\tau}, S_{1:\tau})$, we iterate a fixed window (e.g., $\tau\!=\!12$ steps $\approx\!4$\,s):
\begin{align}
\hat{x}_{\tau+1,\cdot} &= f_\theta(X_{1:\tau}, S_{1:\tau}),\;
\hat{x}_{\tau+2,\cdot}=f_\theta([X_{2:\tau},\hat{x}_{\tau+1,\cdot}], [S_{2:\tau},S_{\tau+1}]), \dots
\end{align}

\subsection{Behavior-from-activity head (PNM)}
\label{sec:pnm}
To map brain states to behavior, we use a permutation-invariant pooling model. Let $X_t\in[0,1]^N$ and $P\in\mathbb{R}^{N\times 3}$. We center/scale $P$, add Fourier features $\psi(p_n)$, and $z$-score spikes per time to remove global gain. Each $[\tilde{x}_{t,n};\psi(p_n)]$ is encoded by a small MLP to $h_{t,n}$; masked mean pooling yields $\bar{h}_t$, a temporal MLP aggregates $\{\bar{h}_{t-\tau+1},\dots,\bar{h}_t\}$, and we predict behavior $y\in\mathbb{R}^{d_\text{beh}}$. 
\emph{Training.} We train one model per behavior dimension with L1 loss ($\ell_1$) on targets in $[0,1]$, AdamW optimizer, linear warm-up (10\% of steps), gradient-clipping at 1.0, and balanced mini-batches that match rare high-magnitude frames to a subset of typical frames for stability.

\subsection{Implementation details and training practice}
\label{sec:impl}

\textbf{Masks/geometry/kernels.} All operations honor neuron masks. Spatial attention uses directional RoPE on positions; temporal attention uses standard RoPE over time \citep{Su2021}. Spatial attention runs with balanced routing and variable-length FlashAttention; temporal attention batches valid neuron rows into a single causal FlashAttention call \citep{Roy2020Routing,Dao2023FlashAttn2}. In practice $w$ is a few hundred, making compute effectively linear in $N$.

\textbf{Numerics.} The model runs end-to-end in \textbf{bf16}; logits and losses are computed in fp32. We enable TF32 matmuls on CUDA and \verb|torch.compile| with dynamic shapes. A CUDA prefetcher overlaps H2D copies and casts spikes/stim to bf16.

\textbf{Optimization.} We use \textbf{MuonWithAuxAdam} \citep{KellerJordanMuon} for matrix/tensor weights ($\geq\!2$D) in the attention \emph{body} and AdamW for gains/biases plus \emph{embed/head}. Typical hyperparameters: Muon learning rate $2\cdot10^{-2}$, AdamW learning rate $5\cdot10^{-4}$ (with betas of $0.9$ and $0.95$), weight decay $10^{-4}$, gradient-clipping at $1.0$. A warmup-cosine schedule (per batch) is applied; we validate several times per epoch and keep the best checkpoint. Random seeds are fixed, and mixed precision plus caching/prefetching keep throughput high.

We use larval zebrafish whole-brain calcium imaging \citep{Chen2018,Ahrens2013,Vladimirov2014} and adopt a subject-level split: models are trained on the training subject fish and evaluated on a single held-out fish reserved for validation. No frames from the held-out subject are used for training or hyperparameter tuning. Unless otherwise noted, all quantitative results and visualizations are computed on the held-out validation fish.

\vspace{0.3em}
\noindent
Together, these design choices yield an interpretable, scalable architecture that maintains single-neuron fidelity while capturing brain-wide interactions and long-range temporal structure.

\section{Results}
\label{sec:results}

\paragraph{Setup.}
We trained and evaluated the Sparse Brain Model (SBM) on brain-wide light-sheet recordings of behaving larval zebrafish, using CASCADE to infer causal spike rates that we convert to spike probabilities for learning \citep{Chen2018,Ahrens2013,Vladimirov2014,Rupprecht2021}. Unless noted otherwise, models use a 4\,s ($\tau{=}12$ steps) context window with teacher forcing at train time and fixed-window autoregression at test time. We want to underscore that even with such a small context window the model is able to predict future neural activity with high fidelity and without predicting collapsing into either over or under activation. 

\subsection{Next-step accuracy, calibration, and rollout stability}

The SBM accurately predicts single-neuron next-step activity while preserving cell-level structure (Fig.~\ref{fig:arch}A--C). In qualitative overlays, predicted spike-probability point clouds (voxelized to a $512{\times}256$ grid and mean-pooled across $z$) closely track ground truth across time, and single-neuron traces show both transient and sustained events being followed (Fig.~\ref{fig:arch}B--C). Quantitatively, next-step mean absolute error is \textbf{$<0.02$} under typical contexts and improves further with longer histories (Fig.~\ref{fig:metrics}B). Predictions are \emph{well calibrated}: the predicted mean spike probabilities closely match the empirical means, indicating reliable uncertainty at the neuron level (Fig.~\ref{fig:metrics}A). During autoregression with a 4\,s sliding window, error accumulation is \emph{sublinear} over horizon length and rollouts remain stable (Fig.~\ref{fig:metrics}C).

\subsection{Distributional fidelity of predicted brain states}

Beyond pointwise error, predicted and real brain states occupy similar low-dimensional manifolds. In PCA and UMAP embeddings, the distributions of next-step predictions largely overlap with those of ground truth, supporting the claim that SBM learns the \emph{distribution} of brain activity rather than overfitting to mean trends \citep{Jolliffe2016,McInnes2018} (Fig.~\ref{fig:metrics}D). This property is important for downstream tasks that depend on realistic population-level coordination rather than isolated spikes.

\subsection{Linking brain state to behavior}

To connect neural dynamics to action, we pair SBM (or ground-truth spikes) with a permutation-invariant behavior head (PNM; Sec.~\ref{sec:pnm}). Using four brain-state time points taken backward in time as input, the PNM predicts behavior on held-out fish with a mean Pearson correlation of \textbf{0.42} (Fig.~\ref{fig:behavior}A). In a behavioral PCA space, predicted behaviors cover much of the distribution of real behaviors, indicating that the neural representations learned by SBM are behaviorally informative \citep{Naumann2016,Jolliffe2016} (Fig.~\ref{fig:behavior}B).

\subsection{Novel behavior generation via gradient-based neural pattern search}
\label{sec:neural_inverse_design}

\begin{figure}
  \centering
  \includegraphics[width=1\linewidth]{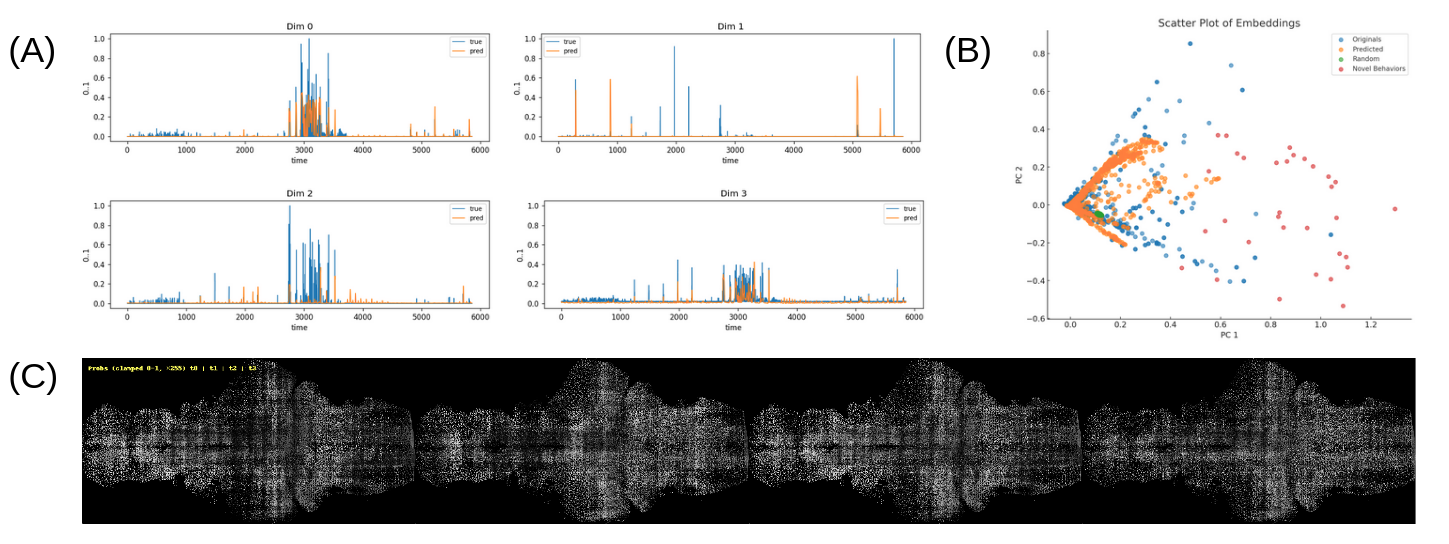}
  \caption{Adding the Peripheral Neural Model (PNM) enables prediction of fish behaviors from brain states—either ground truth or model predictions—and enables the generation of novel behaviors. (A) For a held-out fish, the PNM predicts behaviors from four brain-state time points taken backward in time, achieving a mean Pearson correlation coefficient of 0.42 with ground truth. (B) In PCA space, the PNM replicates a substantial portion of the original behavior distribution. When driven by random neural stimulations, the model generates only a small subset of behaviors, implying that realistic behaviors require highly structured neural patterns; in contrast, optimizing neural inputs via gradient descent to target behaviors yields a much broader distribution of novel behaviors. (C) A visualizes a four-time-point neural state learned by the PNM to elicit a novel behavior.}
  \label{fig:behavior}
\end{figure}

We asked whether brief, structured neural patterns could \emph{generate} new behaviors. A naive baseline—driving the system with \emph{random} neural stimulations—produced only a \emph{small} subset of the behavioral space, collapsing to modes already common in the data (Fig.~\ref{fig:behavior}B, “random”). In contrast, optimizing short sequences of neuron-specific inputs by gradient descent (through the differentiable PNM) discovered \emph{diverse} novel behaviors that populate previously sparse regions of the embedding (Fig.~\ref{fig:behavior}B, “Novel Behaviors”).

Together, these results show that (i) the SBM provides calibrated, accurate forecasts at single-neuron resolution with stable rollouts, (ii) predicted brain states retain distributional fidelity, and, (iii) when we coupled the SBM to the PNM, \emph{neural inverse design} enables gradient-learned patterns substantially expand reachable behavior beyond what random stimulation achieves.

\section{Discussion}

Current practice trades off scale, interpretability, and behavioral grounding. POCO-style pipelines are not spike rate native and are not conditioned on sensory information \citep{Duan2025POCO}. Video U-Net approaches demand multiple large GPUs and rely on local convolutions that cannot flexibly attend between distant neurons \citep{Ronneberger2015,Cicek2016}, and state-of-the-art video diffusion systems still inherit heavy convolutional backbones \citep{Ho2022VDM,Ho2022ImagenVideo}. These constraints hinder models that must predict single-neuron activity at whole-brain scale while remaining faithful to the distribution of brain states and usable for behaviorally relevant analysis.

Our results show that sparse attention can meet these requirements by separating spatial from temporal reasoning and by operating directly on spike probabilities inferred from calcium imaging \citep{Rupprecht2021}. The dynamic connectome layer provides interpretable neuron–neuron influences, while the temporal layer captures within-cell dynamics with causal self-attention informed by rotary position embeddings \citep{Vaswani2017,Su2021}. Fidelity in PCA and UMAP spaces indicates that the model preserves the structure of the brain’s activity distribution rather than overfitting to pointwise errors \citep{Jolliffe2016,McInnes2018}. The Peripheral Neural Model links predicted brain states to behavior and enables gradient-based synthesis of neural patterns that broaden the repertoire of achievable actions, creating an efficient path for rapid in silico experiments that can guide targeted in vivo tests \citep{Naumann2016}.

If successful, this framework can make in-silico experimentation routine: researchers could screen perturbations, prioritize targets, and design closed-loop interventions that elicit desired behaviors before committing to in vivo trials. Interpretable sparse attention may provide a bridge between data-driven prediction and mechanistic theory by exposing testable neuron-to-neuron influences, enabling integration with anatomical connectomes and cell-type maps, and guiding targeted stimulation or pharmacological interventions. Extending the approach to richer sensory contexts, longer timescales, and other species could yield general-purpose whole-brain foundation models that support hypothesis generation for complex states such as sleep and learning, improve the data efficiency of experimental programs, and ultimately enable principled control of neural circuits.

% \begin{ack}
% Use unnumbered first level headings for the acknowledgments. All acknowledgments

% Do {\bf not} include this section in the anonymized submission, only in the final paper. You can use the \texttt{ack} environment provided in the style file to automatically hide this section in the anonymized submission.
% \end{ack}

% \section*{References}

% References follow the acknowledgments in the camera-ready paper. Use unnumbered first-level heading for
% the references. Any choice of citation style is acceptable as long as you are
% consistent. It is permissible to reduce the font size to \verb+small+ (9 point)
% when listing the references.
% Note that the Reference section does not count towards the page limit.
\medskip

% {
% \small

% [1] Alexander, J.A.\ \& Mozer, M.C.\ (1995) Template-based algorithms for
% connectionist rule extraction. In G.\ Tesauro, D.S.\ Touretzky and T.K.\ Leen
% (eds.), {\it Advances in Neural Information Processing Systems 7},
% pp.\ 609--616. Cambridge, MA: MIT Press.

% [2] Bower, J.M.\ \& Beeman, D.\ (1995) {\it The Book of GENESIS: Exploring
%   Realistic Neural Models with the GEneral NEural SImulation System.}  New York:
% TELOS/Springer--Verlag.

% [3] Hasselmo, M.E., Schnell, E.\ \& Barkai, E.\ (1995) Dynamics of learning and
% recall at excitatory recurrent synapses and cholinergic modulation in rat
% hippocampal region CA3. {\it Journal of Neuroscience} {\bf 15}(7):5249-5262.
% }
\bibliographystyle{plainnat}
\bibliography{refs}

%%%%%%%%%%%%%%%%%%%%%%%%%%%%%%%%%%%%%%%%%%%%%%%%%%%%%%%%%%%%

% \appendix

% \section{Technical Appendices and Supplementary Material}
% Technical appendices with additional results, figures, graphs and proofs may be submitted with the paper submission before the full submission deadline (see above), or as a separate PDF in the ZIP file below before the supplementary material deadline. There is no page limit for the technical appendices.

%%%%%%%%%%%%%%%%%%%%%%%%%%%%%%%%%%%%%%%%%%%%%%%%%%%%%%%%%%%%

\end{document}